\documentclass[english]{article}
\usepackage{lmodern}
\usepackage[T1]{fontenc}
\usepackage[latin9]{inputenc}
\usepackage{amssymb}
\usepackage{graphicx}
\usepackage{esint}
\usepackage{babel}
\makeatletter
\newcommand{\lyxaddress}[1]{
\par {\centering #1
\vspace{1.4em}
\noindent\par}
}

\newcommand{\lyxemail}[1]{
\par {\centering #1
\vspace{1.4em}
\noindent\par}
}
\makeatother

\usepackage{babel}
\begin{document}

\title{Thermodynamics of the universe bounded by the cosmological event horizon
 and dominated by the tachyon fluid}
\author{Fei-Quan Tu, Yi-Xin Chen\footnote{Corresponding author}}
\date{}
\maketitle

\lyxaddress{Zhejiang Institute of Modern Physics, \\Zhejiang University, Hangzhou, 310027, P. R. China}
\lyxemail{Email: fqtuzju@foxmail.com, yxchen@zimp.zju.edu.cn}

\begin{abstract}
Our aim is to investigate the thermodynamic properties of the universe bounded
by the cosmological event horizon and dominated by the
tachyon fluid. We give two different
laws of evolution of our universe. Further, we show the first law and the generalized second law of
thermodynamics (GSLT) are both satisfied in two cases, but their properties of the
thermodynamic equilibrium
are totally different. Besides, under our solutions, we find the validity of the laws of
thermodynamics is irrelevant with the parameters of the tachyon fluid.
Finally, we conclude that the universe bounded by the cosmological event horizon
and dominated by the tachyon fluid has a good thermodynamic description.
In turn, the thermodynamic description can provide a good physical interpretation
for the dynamic evolution of our universe due to the equivalence
between the first law of thermodynamics and the Friedmann equation to some extent.

\end{abstract}

\section*{1. Introduction}

Numerous astro-observations tell us that our universe is in accelerating
expansion at present\cite{key-1,key-2}. The most common way to describe
the effects of the current accelerating expansion is to introduce
a cosmological constant in the Einstein's equation. This model, called
$\Lambda-$model, well explains the astro-observations. According to
this model, about $73\%$ of the total energy in our universe is the
dark energy. However, the origin of the cosmological constant (or called the
dark energy) is unknown. In 2002, Sen\cite{key-3,key-4}showed that
the classical decay of unstable D-branes in bosonic and superstring
theories produces pressureless gas with non-zero energy density. Further
he gave a description of this phenomenon in an effective field theory.
Afterwards, the idea was applied to cosmology by Gibbons\cite{key-5},
Padmanabhan\cite{key-6}, Feinstein\cite{key-7}, etc. The effective
action of the tachyon field associated with the decay of unstable
D-branes was used as the fundamental action to investigate the inflation
(see, for example, \cite{key-7,key-8,key-9,key-10}) or the current
accelerating expansion (see, for instance, \cite{key-6,key-11,key-12,key-13,key-14}).
This way to study cosmology is attractive because either the dark energy which causes the current
accelerating expansion or the inflation potential originates from a fundamental theory.
Therefore, the tachyon field is a good candidate of the dark energy.

On the other hand, the feature which the field equation doesn't have
any direct physical interpretation and the lack of an elegant principle
which can lead to the dynamics of gravity are unsatisfied as have
been pointed out by Padmanabhan. He proposed that an interpretation
should be based on the thermodynamics of horizons\cite{key-15}. Fortunately,
since the black hole thermodynamics\cite{key-16,key-17} was discovered
in the 1970's, the relationship between thermodynamics and the horizon
of black hole has been generally accepted. Furthermore, the Einstein
equation can be derived from the proportionality of entropy and horizon
area together with the relation $\delta Q=TdS$ connecting heat, entropy
and temperature\cite{key-18}. The thermodynamic interpretation for
the field equations in any diffeomorphism invariant theory of the
gravity was also provided\cite{key-15}. The Friedmann equation with spatial
curvature can also be derived from the first law of thermodynamics
on the apparent horizon of the FRW universe\cite{key-19}. Hence,
there exists a deep connection between the cosmological horizon and
thermodynamics, and it is important and significant to investigate
the thermodynamic properties of cosmological horizon. Besides, the
thermal properties of dark energy have also been generally discussed,
for example, see Ref.\cite{key-20,key-21,key-22,key-23}.

In a cosmological model with a positive cosmological constant, the
cosmological event horizon is the boundary of the past of the observer's
world line. This horizon is similar to the black-hole event horizon
which can be described by thermodynamics\cite{key-24}. There exists
an event horizon for our universe which is in accelerating expansion,
so it is important and interesting to investigate the first law of thermodynamics
and the GSLT of the universe bounded by the cosmological event horizon
and dominated by the tachyon fluid (in this paper we should
use terms like the tachyon field and the tachyon fluid interchangeably).

If the usual form of Hawking temperature $T=1/(2\pi R)$ where $R$ is the radius of the horizon
is used to investigate the cosmological event horizon,
then the conclusion that the cosmological event horizon is unphysical from the point of view of the laws
of thermodynamics is obtained\cite{key-25}. However, if the modified Hawking temperature is taken,
the correct laws of thermodynamics on the cosmological event horizon can be obtained\cite{key-26}.
We will study the laws of thermodynamics of the universe bounded by the cosmological event horizon
and dominated by the tachyon field through using the modified temperature in this paper. First,
the tachyon field which only changes with time is taken as
the form $\phi=At^{\beta}$ where $A$ and $\beta$ are positive
constants in the flat FRW universe. By substituting this form into
the Friedmann equation and the continuity equation (this is also the equation of motion of
the tachyon field $\phi$), we obtain the
laws of growth of the scale factor of the universe. Then we restrict the
constant $\beta$ to make it satisfy $\frac{1}{2}<\beta\leq1$ by
comparing the results of the theoretical calculation with that of
the astronomical observations. In the case of $\beta=1$, the tachyon potential
is proportional to inverse square of the tachyon field and we get
the restriction $0<A<\frac{\sqrt{6}}{3}$ because our universe is
in accelerating expansion. In the case of $\frac{1}{2}<\beta<1$
, $\beta=\frac{2}{3}$ is taken due to the similarity of the properties
of evolution of the universe and $A=1$ is chosen in order to
see the properties of evolution of the universe more clearly.
By investigating the above two cases, we show
that they both satisfy the first law of thermodynamics and the GSLT,
but their properties of the thermodynamic equilibrium
are totally different. The universe dominated by the field $\phi=At$
can't reach the thermodynamic equilibrium while the universe dominated
by the field $\phi=At^\beta$ where $\frac{1}{2}<\beta<1$ can reach the thermodynamic equilibrium.
Besides,  under our solutions, we find the validity of the laws
of thermodynamics is irrelevant with the parameters of the tachyon fluid.
In the paper the units are chosen with $\ensuremath{c=\hbar=k_{B}=1}$ and
the signature of the spacetime metric is taken as $(+,-,-,-)$.

\section*{2. Properties of evolution of the universe dominated by the tachyon
fluid}

The effective Lagrangian density of the tachyon field in string theory
is taken as\cite{key-5}
\begin{equation}
L=-V(\phi)\sqrt{1-g^{\mu\nu}\partial_{\mu}\phi\partial_{\nu}\phi},
\end{equation}
where $\phi$ is the tachyon field and $V(\phi)$ is the tachyon potential.

Thus, the action of coupling this tachyon field with gravity is
\begin{equation}
S=\int d^{4}x\sqrt{-g}\left(\frac{R}{16\pi G}+L\right),
\end{equation}
where $R$ is the scalar curvature. Varying the action (2) with respect to the metric $g^{\mu\nu}$,
we obtain the Einstein equation
\begin{equation}
R_{\mu\nu}-\frac{1}{2}g_{\mu\nu}R=8\pi GT_{\mu\nu},
\end{equation}
where the stress tensor is defined as
\begin{equation}
T_{\mu\nu}=\frac{2}{\sqrt{-g}}\frac{\partial\left(\sqrt{-g}L\right)}{\partial g^{\mu\nu}}=\left(\frac{\partial_{\mu}\phi\partial_{\nu}\phi}{\sqrt{1-\partial_{\lambda}\phi\partial^{\lambda}\phi}}
+g_{\mu\nu}\sqrt{1-\partial_{\lambda}\phi\partial^{\lambda}\phi}\right)V(\phi)
\end{equation}
for the tachyon field. For the perfect fluid, its form is
\begin{equation}
T_{\mu\nu}=(\rho+p)u_{\mu}u_{\nu}-pg_{\mu\nu}.
\end{equation}
If the tachyon field is considered as an effective fluid, then we
can obtain the corresponding energy density, pressure and velocity
\begin{equation}
\rho=\frac{V(\phi)}{\sqrt{1-\partial_{\lambda}\phi\partial^{\lambda}\phi}},\qquad p=-V(\phi)\sqrt{1-\partial_{\lambda}\phi\partial^{\lambda}\phi},\qquad u_{\mu}=\frac{\partial_{\mu}\phi}{\sqrt{1-\partial_{\lambda}\phi\partial^{\lambda}\phi}}.
\end{equation}

In the flat FRW universe, the metric is
\begin{equation}
ds^{2}=-dt^{2}+a^{2}(t)(dx^{2}+dy^{2}+dz^{2}),
\end{equation}
where $a(t)$ is the scale factor. From the metric, we know the tachyon
field is irrelevant with the space coordinates. The Friedmann equation
is
\begin{equation}
H^{2}=\frac{8\pi G}{3}\rho=\frac{8\pi G}{3}\frac{V(\phi)}{\sqrt{1-\dot{\phi}}},
\end{equation}
where a dot refers to a derivative with respect to the cosmic time and $H=\frac{\dot{a}(t)}{a(t)}$ is the Hubble constant.

Varying the action (2) with respect to the tachyon field $\phi$, we obtain the following equation
of motion
\begin{equation}
\frac{\ddot{\phi}}{1-\dot{\phi}^{2}}+3H\phi+\frac{\dot{V}}{V\dot{\phi}}=0.
\end{equation}
This equation is equivalent to the continuity
equation which is
\begin{equation}
\dot{\rho}+3H(\rho+p)=0.
\end{equation}
Combining Eq.(8) with Eq.(10), we obtain
\begin{equation}
\dot{H}=-\frac{3}{2}\dot{\phi}^{2}H^{2}.
\end{equation}

Because the specific potential form for the tachyon field isn't fixed by the
string theory, we can investigate the different form to see the properties of evolution
of the universe. In this paper, the tachyon field $\phi$ is taken as the following form
\begin{equation}
\phi=At^{\beta},
\end{equation}
where $A$ and $\beta$ are positive constants.

If $\beta=1$, we get the solution of Eq.(11)
\begin{equation}
a(t)=t^{\frac{2}{3A^{2}}},
\end{equation}
where we have assumed $H(0)=\infty$. From this form, we know the
universe is in accelerating expansion when $A^{2}<\frac{2}{3}$. So
the parameter $A$ has the restriction $0<A<\frac{\sqrt{6}}{3}$ when
the field is $\phi=At$. Then the energy density, the potential and
the pressure are respectively read as
\begin{equation}
\rho=\frac{1}{6\pi GA^{4}t^{2}},
\end{equation}
\begin{equation}
V(t)=\frac{\sqrt{1-A^{2}}}{6\pi GA^{4}t^{2}}
\end{equation}
and
\begin{equation}
p=\frac{A^{2}-1}{6\pi GA^{4}t^{2}}.
\end{equation}
Further we can obtain the cosmological event horizon
\begin{equation}
R_{E}=a(t)\int_{t}^{\infty}\frac{dt^{\prime}}{a(t^{\prime})}=\frac{3A^{2}}{2-3A^{2}}t.
\end{equation}

If $\beta\neq1$, we get the solution of Eq.(11)
\begin{equation}
a(t)=exp\left(\frac{(2\beta-1)t^{2-2\beta}}{3A^{2}\beta^{2}(1-\beta)}\right),
\end{equation}
where we have taken the initial values $a(0)=1$ and $H(0)=\infty$. We obtain the Hubble constant
\begin{equation}
H=\frac{2(2\beta-1)}{3A^{2}\beta^{2}}t^{1-2\beta}.
\end{equation}
From this form, we know the universe is in accelerating expansion
when $\frac{1}{2}<\beta<1$ and $\beta>1$.

The properties of evolution of the universe when $\beta=\frac{2}{3}$
are similar to that of $\frac{1}{2}<\beta<1$ while the properties of
evolution of the universe when $\beta=2$ are similar to that of
$\beta>1$, so we take the specific values $A=1$, $\beta=\frac{2}{3}$ and $\beta=2$
in order to see the properties of evolution more clearly.

When $A=1$ and $\beta=\frac{2}{3}$ , we have the scale factor
\begin{equation}
a(t)=exp(\frac{3}{4}t^{2/3}).
\end{equation}
Then we obtain the Hubble constant
\begin{equation}
H=\frac{1}{2t^{1/3}},
\end{equation}
and the energy density
\begin{equation}
\rho=\frac{3}{32\pi Gt^{2/3}}.
\end{equation}
The potential and the pressure are respectively
\begin{equation}
V(t)=\sqrt{1-\frac{4}{9t^{2/3}}}\frac{3}{32\pi Gt^{2/3}}
\end{equation}
and
\begin{equation}
p=(\frac{4}{9t^{2/3}}-1)\frac{3}{32\pi Gt^{2/3}}.
\end{equation}
Moreover, the cosmological event horizon is
\begin{equation}
R_{E}=a(t)\int_{t}^{\infty}\frac{dt^{\prime}}{a(t^{\prime})}=\exp\left(\frac{3}{4}t^{2/3}\right)\frac{2}{3}\left[3\exp\left(-\frac{3}{4}t^{2/3}\right)t^{1/3}+\sqrt{3\pi}erfc\left(\frac{\sqrt{3}}{2}t^{1/3}\right)\right],
\end{equation}
where  $erfc(x)$ is the error function, its definition
is $erfc(x)=\frac{2}{\sqrt{\pi}}\int_{x}^{\infty}e^{-t^{2}}dt$.

When $A=1$ and $\beta=2$, we have the scale factor
\begin{equation}
a(t)=exp(-\frac{1}{4t^{2}}).
\end{equation}
Then we obtain the potential
\begin{equation}
V(t)=\sqrt{1-4t^{2}}\frac{3}{32\pi Gt^{6}}.
\end{equation}
From the above potential, we know the time $t$ should satisfy $0<t<\frac{1}{2}$,
so the cosmological horizon doesn't exist. Besides, the rate of
the expansion is so slow that it isn't consistent with the result of the astronomical observation.
Hence we should discard such solution.

In conclusion, we obtain the solution $\frac{1}{2}<\beta\leq1$ when
we take the form of the field $\phi(t)=At^{\beta}$. Regardless of $\beta=1$ or
$\frac{1}{2}<\beta<1$, the law of evolution of the universe can explain
the inflation in the early universe or the current
accelerating expansion of the universe when the appropriate positive constants are taken.
Here, we would like to point out that  the author in Ref[7]
assumed the form of scale factor of the universe to study the potential form of the tachyon field.
However, we take the solutions of the equation of motion of the tachyon field
to investigate the evolution of the universe. Our way to the results are more natural. Moreover, we can assume
the tachyon field has the other form and analyze the relevant solutions,
but the process is analogous.

\section*{3. Thermodynamics of the universe dominated by the tachyon fluid}
In this section, our main aim is to investigate the first and generalized second
law of thermodynamics for the universe bounded by the cosmological event horizon and
dominated by the tachyon fluid $\phi$ whose forms are taken as that of the solutions in the above section.

\subsection*{3.1. First law of thermodynamics of the universe dominated by the tachyon fluid}
\label{sec:intro}

In the homogenous and isotropic universe, the metric can be expressed
as\cite{key-27}
\begin{equation}
ds^{2}=h_{ij}dx^{i}dx^{j}+R^{2}d\Omega_{2}^{2},
\end{equation}
where $i$, $j$ can take values $0$ and $1$, $R=a(t)r$
and the 2-dimensional metric $h_{ab}=diag(-1,a^{2}/(1-\kappa r^{2}))$
in which $\kappa$ is the spatial curvature constant. A scalar quantity
is defined as
\begin{equation}
\chi=h^{ij}\partial_{i}R\partial_{j}R.
\end{equation}
If the scalar quantity $\chi=0$, it gives $R_{A}=\frac{1}{\sqrt{H^{2}+\frac{\kappa}{a^{2}}}}$
which is called apparent horizon. The surface gravity on the apparent
horizon is defined as
\begin{equation}
\kappa_{A}=-\frac{1}{2}\frac{\partial\chi}{\partial R}\mid_{R=R_{A}}=\frac{1}{R_{A}}
\end{equation}
and the corresponding Hawking temperature is defined as
\begin{equation}
T_{A}=\frac{|\kappa_{A}|}{2\pi}=\frac{1}{2\pi R_{A}}.
\end{equation}
Based on Ref.\cite{key-26}, the Hawking temperature on the cosmological
event horizon $R_{E}$ is defined as
\begin{equation}
T_{E}=\frac{|\kappa_{E}|}{2\pi}=-\frac{1}{4\pi}\frac{\partial\chi}{\partial R}\mid_{R=R_{E}}=\frac{R_{E}}{2\pi R_{A}^{2}}.
\end{equation}

In this paper, we take the flat FRW metric (7), so the apparent horizon
$R_{A}=\frac{1}{H}.$ For the tachyon field which is $\phi=At$, the
scale factor grows with time as Eq.(13) and the cosmological event
horizon is Eq.(17). The amount of energy
flux across the horizon during the time interval $dt$ is\cite{key-28}
\begin{equation}
-dE_{H}=4\pi R_{H}^{2}T_{\mu\nu}k^{\mu}k^{\nu}dt
\end{equation}
with $k^{\mu}=\frac{1}{\sqrt{v}}(v,-v,0,0)$ a null vector. Hence,
for the cosmological horizon, we obtain
\begin{equation}
-dE=4\pi R_{E}^{3}H(\rho+p)dt=\frac{12A^{2}dt}{\left(2-3A^{2}\right)^{3}G},
\end{equation}
where we have used the relation $v=HR_{E}$. According to the area-entropy
relation $S_{E}=\frac{\pi R_{E}^{2}}{G}$, we get
\begin{equation}
T_{E}dS_{E}=\frac{12A^{2}}{(2-3A^{2})^{3}G}dt.
\end{equation}
Thus we see the first law of thermodynamics $-dE=TdS$ is kept on
the cosmological event horizon when the tachyon field is taken as
$\phi=At$.

For the tachyon field  $\phi=t^{\frac{2}{3}}$, the scale
factor grows with time as Eq.(20) and the cosmological event horizon
is Eq.(25). Compared to the term $3\exp\left(-\frac{3}{4}t^{2/3}\right)t^{1/3}$,
the error function term $\sqrt{3\pi}erfc\left(\frac{\sqrt{3}}{2}t^{1/3}\right)$
is very small when $t$ is relatively large (for example $t>10$), so we
can discard it. Thus the cosmological event horizon is
\begin{equation}
R_{E}\approx2t^{1/3}.
\end{equation}
The amount of energy flux across the cosmological event horizon is
\begin{equation}
-dE=4\pi R_{E}^{3}H(\rho+p)dt=\frac{2dt}{3Gt^{2/3}}.
\end{equation}
On the other hand, we can obtain

\begin{equation}
T_{E}dS_{E}=\frac{2dt}{3Gt^{2/3}}.
\end{equation}
Hence the first law of thermodynamics $-dE=TdS$ is approximately
satisfied on the cosmological event horizon when the tachyon field is taken
as $\phi=t^{\frac{2}{3}}$. Due to the similarity of the properties,
the first law of thermodynamics should be correct on the cosmological
event horizon when the tachyon field is $\phi=At^{\beta}$ where
the parameter $\beta$ satisfies $\frac{1}{2}<\beta<1$.

In this subsection, we show that the first law of thermodynamics is correct
or approximately correct on the cosmological event horizon in the flat FRW
universe dominated by the tachyon field whose rate of change is
$\phi=At^{\beta}(\frac{1}{2}<\beta\leq1)$. Furthermore, the
first law of the thermodynamics derives from the Friedmann equation,
so they are equivalent to some extent.

\subsection*{3.2. GSLT and thermodynamic equilibrium of the universe dominated by the tachyon fluid}
\label{sec:intro}

In this subsection, our main aim is to investigate whether the universe
bounded by the cosmological event horizon and dominated by the tachyon fluid satisfies
the GSLT and approaches an thermodynamic equilibrium or not.
By means of the entropy functions, the GSLT and the thermodynamic
equilibrium configuration can be expressed as\cite{key-29,key-30}
\begin{equation}
(1)\quad\dot{S}_{h}+\dot{S}_{fh}\geq0,\qquad(2)\quad\ddot{S}_{h}+\ddot{S}_{fh}<0
\end{equation}
where $S_{h}$ and $S_{fh}$ represent the entropies of the horizon
and the tachyon fluid respectively.

By using some thermodynamic relations and the assumption which the temperature of the fluid
is same as that of the horizon, the relations
\begin{equation}
\dot{S}_{E}+\dot{S}_{fE}=\frac{8\pi^{2}R_{E}(\rho+p)}{H}\left(R_{E}-\frac{1}{H}\right)
\end{equation}
and
\begin{equation}
\ddot{S}_{E}+\ddot{S}_{fE}=8\pi^{2}(\rho+p)\left(R_{E}-\frac{1}{H}\right)
\left[-\left\{ \frac{R_{E}}{2}\left(1-3\frac{p}{\rho}\right)+\frac{1}{H}+3R_{E}\frac{\dot{p}}{\dot{\rho}}\right\} +R_{E}\left\{ 1-\frac{3(1+\frac{p}{\rho})}{2(HR_{E}-1)}\right\} \right]
\end{equation}
were obtained in Ref.\cite{key-30}.

Now, we analyze the GSLT and the thermodynamic equilibrium for the universe dominated by the following
different potential of the tachyon fluid:

1. The potential $V(t)=\frac{\sqrt{1-A^{2}}}{6\pi GA^{4}t^{2}}.$
For this potential, our universe evolutes as Eq.(13). For entropy variations, Eq.(40) and
Eq.(41) become
\begin{equation}
\dot{S}_{E}+\dot{S}_{fE}=\frac{8\pi^{2}R_{E}A^{2}\rho}{H}\frac{9A^{4}t}{2(2-3A^{2})}
\end{equation}
and
\begin{equation}
\ddot{S}_{E}+\ddot{S}_{fE}=8\pi^{2}A^{2}\rho\left(R_{E}-\frac{1}{H}\right)^{2}.
\end{equation}
From the Eq.(42), we know $\dot{S}_{E}+\dot{S}_{fE}>0$ because of
$A^{2}<\frac{2}{3}$, which indicates that the total entropy is increasing,
so the GSLT is satisfied. However, $\ddot{S}_{E}+\ddot{S}_{fE}>0$
which indicates our universe can't reach the thermodynamic equilibrium from the Eq.(43).

2. The potential $V(t)=\sqrt{1-4t^{2}}\frac{3}{32\pi Gt^{6}}$. For
this potential, our universe evolutes as Eq.(20). For entropy variations, Eq.(40) and Eq.(41)
become
\begin{equation}
\dot{S}_{E}+\dot{S}_{fE}=\frac{32\pi^{2}R_{E}\rho}{9Ht^{2/3}}\left(R_{E}-\frac{1}{H}\right)
\end{equation}
and
\begin{equation}
\ddot{S}_{E}+\ddot{S}_{fE}=\frac{32\pi\rho}{9t^{2/3}}R_{E}\left(R_{E}-\frac{1}{H}\right)\left[F(t)- \frac{2}{3(HR_{E}-1)t^{2/3}}\right]
\end{equation}
where $F(t)=1-\left\{ \frac{1}{2}\left(1-3\frac{p}{\rho}\right)+\frac{1}{HR_{E}}+3\frac{\dot{p}}{\dot{\rho}}\right\}$.
From the Eq.(44), we know $\dot{S}_{E}+\dot{S}_{fE}\geq0$ because
$R_{E}-\frac{1}{H}\geq0$.
This indicates that the total entropy is increasing, so the GSLT is satisfied.
Further, we see $\dot{S}_{E}+\dot{S}_{fE}=0$ when $t\rightarrow\infty$.
From the Figure 1, we know that $\ddot{S}_{E}+\ddot{S}_{fE}\leq0$ which indicates that our universe
is approaching a thermodynamic equilibrium configuration.

\begin{figure}
\centering
\includegraphics[width=8cm]{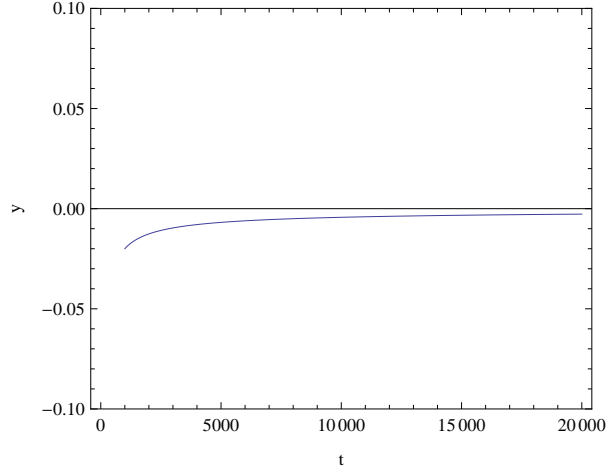}
\caption{In the picture, $y$ represents $F(t)- \frac{2}{3(HR_{E}-1)t^{2/3}}$. We can know $y<0$ and $y$ tends to $0$
when time $t$ is large.}
\end{figure}

Analyzing the GSLT and the thermodynamic equilibrium configuration for the universe
dominated by the above different potential, we show they both satisfy
the GSLT, but their properties of the thermodynamic equilibrium are totally
different. The universe dominated by the potential $V(t)=\frac{\sqrt{1-A^{2}}}{6\pi GA^{4}t^{2}}$
can't reach the thermodynamic equilibrium while the universe dominated
by the potential $V(t)=\sqrt{1-4t^{2}}\frac{3}{32\pi Gt^{6}}$ can
reach the thermodynamic equilibrium.

.

\section*{4. Conclusion and discussion}

Thermodynamics of the universe bounded by the cosmological event horizon
has been investigated in the perfect fluid dark model (that is, the equation of state of the dark energy is
$p=\omega\rho$ where $\omega<-\frac{1}{3}$ and $\omega$ is a constant) and in the holographic
dark energy model\cite{key-26,key-31}. The thermodynamical properties
related to the apparent horizon and the cosmological event horizon were studied
in Ref.\cite{key-25} where the dark energy was also taken as holographic
fluid. However, all the cosmological constant, the perfect fluid dark energy
and the holographic dark energy don't have a basic physical interpretation.
In this paper, the tachyon field is taken as the fluid of dark energy
in our universe. As have been pointed out in introduction, the tachyon
field originates from the decay of unstable D-branes in string theory,
so it has a good physical origin. This is our main motivation to choose
the tachyon fluid as the dark energy. Besides, a large number
of authors have studied the field as the dark energy which causes
the current accelerating expansion or the inflation field and their
conclusions showed the tachyon field is a good candidate of the dark
energy. As we have known, the cosmological event horizon is the boundary
of the past of the observer's world line. On the other hand, the horizon
of the black hole is the boundary of observable area for the outside
observers. Hence the properties of the cosmological event horizon
is similar to that of the black hole. Now that the horizon of the
black hole can be described by the laws of thermodynamics, then the
cosmological event horizon should also be described by thermodynamics.
Therefore, the cosmological event horizon has a better physical interpretation
than the Hubble horizon or the apparent horizon under the description of
thermodynamics of the cosmological horizon. This is the main reason that
we choose the cosmological event horizon as the boundary of the universe in this paper.

First, the tachyon field which changes with time is taken as the
form $\phi=At^{\beta}$ in the flat FRW universe.
Then, by combining the Friedmann equation with the continuity equation
and analyzing the laws of growth of the scale factor, we find that
$\frac{1}{2}<\beta\leq1$ is a good interval to describe the effects
of accelerating expansion. The law of growth of the scale factor is
$a(t)=t^{\frac{2}{3A^{2}}}$ with $0<A<\frac{\sqrt{6}}{3}$ when $\beta=1$,
whose form is similar to that of standard cosmological model. While,
the scale factor is $a(t)=exp\left(\frac{(2\beta-1)t^{2-2\beta}}{3\beta^{2}(1-\beta)A^2}\right)$
when $\frac{1}{2}<\beta<1$. However, Regardless of $\beta=1$ or
$\frac{1}{2}<\beta<1$, the law of evolution of the universe can explain
the inflation in the early universe or the current
accelerating expansion of the universe when the appropriate positive constants are taken.
In the case of $\beta=1$, the first law of thermodynamics and the GSLT are
satisfied but the thermodynamic equilibrium can't be reached. On the
other hand, in the case of $\frac{1}{2}<\beta<1$, the first law of thermodynamics
is approximately correct ( because the error function can be not considered
when $t$ is relatively large), the GSLT are satisfied and the thermodynamic
equilibrium can be reached.
Besides, under our solutions, we find the validity of
the laws of thermodynamics is irrelevant with the parameters of the tachyon field.
Finally, we can conclude the universe bounded by the cosmological event horizon and dominated by the tachyon
fluid has a good thermodynamic description. In turn, the thermodynamic description
can provide a good physical interpretation for the dynamic evolution of our universe due to the equivalence
between the first law of thermodynamics and the Friedmann equation to some extent.

\section*{Acknowledgments}

This work is supported in part by the NSF of China Grant No. 11075138 and No. 11375150.

\end{document}